\documentclass{article}

\usepackage{arxiv}          
\usepackage[utf8]{inputenc} 
\usepackage[T1]{fontenc}    
\usepackage{hyperref}       
\usepackage{url}            
\usepackage{booktabs}       
\usepackage{amsfonts}       
\usepackage{nicefrac}       
\usepackage{microtype}      
\usepackage{amsmath}        
\usepackage{amsthm}         
\usepackage{enumitem}       
\usepackage[normalem]{ulem} 
\usepackage{graphicx}
\usepackage{array}
\usepackage{lscape}
\usepackage{natbib}
\usepackage{doi}
\usepackage{placeins}
\usepackage{color}

\title{A Flexible Approach for Predictive Biomarker Discovery}

\date{}

\author{
  Philippe Boileau \\
  Graduate Group in Biostatistics and\\
  Center for Computational Biology,\\
  University of California, Berkeley, and\\
  \texttt{philippe\_boileau@berkeley.edu} \\
  \And
  Nina Ting Qi \\
  Genentech Inc. \\
  \texttt{qit3@gene.com} \\
  \And
  Mark van der Laan \\
  Division of Biostatistics, \\
  Department of Statistics, and\\
  Center for Computational Biology,\\
  University of California, Berkeley\\
  \texttt{laan@berkeley.edu} \\
  \And
  Sandrine Dudoit \\
  Department of Statistics, \\
  Division of Biostatistics, and \\
  Center for Computational Biology,\\
  University of California, Berkeley\\
  \texttt{sandrine@stat.berkeley.edu} \\
  \And
  Ning Leng \\
  Genentech Inc. \\
  \texttt{lengn@gene.com} \\
}

\renewcommand{\headeright}{Preprint}
\renewcommand{\undertitle}{Preprint}

\newtheorem{theorem}{Theorem}
\newtheorem{corollary}{Corollary}

\newtheoremstyle{assumption}%
  {}{}%
  {\itshape}{}%
  {\bfseries}{}%
  { }%
  {(\thmname{#1}\thmnumber{#2})}

\theoremstyle{assumption}
\newtheorem{assumption}{A}

\newcommand{\sd}[1]{}
\newcommand{\pb}[1]{}
\newcommand{\nl}[1]{}

\begin{document}
\maketitle

\begin{abstract}
  An endeavor central to precision medicine is predictive biomarker discovery;
  they define patient sub-populations which stand to benefit most, or least,
  from a given treatment. The identification of these biomarkers is
  often the byproduct of the related but fundamentally different task of
  treatment rule estimation. Using treatment rule estimation methods to
  identify predictive biomarkers in clinical trials where the number of
  covariates exceeds the number of participants often results in high false
  discovery rates. The higher than expected number of false positives
  translates to wasted resources when conducting follow-up experiments for drug
  target identification and diagnostic assay development. Patient outcomes are
  in turn negatively affected. We propose a variable importance parameter for
  directly assessing the importance of potentially predictive biomarkers, and
  develop a flexible nonparametric inference procedure for this estimand. We
  prove that our estimator is double-robust and asymptotically linear under
  loose conditions on the data-generating process, permitting valid inference
  about the importance metric. The statistical guarantees of the method are
  verified in a thorough simulation study representative of randomized control
  trials with moderate and high-dimensional covariate vectors. Our procedure is
  then used to discover predictive biomarkers from among the tumor gene
  expression data of metastatic renal cell carcinoma patients enrolled in
  recently completed clinical trials. We find that our approach more readily
  discerns predictive from non-predictive biomarkers than procedures whose
  primary purpose is treatment rule estimation. An open-source software
  implementation of the methodology, the uniCATE R package, is briefly
  introduced.
\end{abstract}


\section{Introduction}

Precision medicine is now a chief focus of the biomedical establishment. Its
promise of tailored interventions and therapies is impossible to overlook,
potentially spelling major improvements in patient outcomes
\citep{kraus2018,ginsburg2018}. Much effort has therefore been invested in the
development of quantitative methods capable of uncovering patient
sub-populations which benefit more, or less, from novel therapies than the
standard of care.

These groups of patients are distinguished from one another based upon diverse
biometric measurements referred to as predictive biomarkers
\citep{royston2008,kraus2018}. Examples include age, sex at birth, ethnicity,
and gene expression data taken from various tissue samples. Once identified,
these biomarkers may provide clinicians and biologists with mechanistic insight
about the disease or therapy, and spur the development of diagnostic tools for
targeted treatment regimes.

The statistical discovery of predictive biomarkers has, to date, largely been a
byproduct of conditional average treatment effect (CATE) estimation. This
typically unknown parameter contrasts the expected outcomes of patients under
different treatments as a function of their characteristics, thereby defining
the optimal treatment rule. Employing an estimate of the CATE, clinicians can
identify a subgroup of patients that draws most benefit from a therapy. When
estimated using sparse modelling or otherwise interpretable methods,
interpretable machine learning algorithms can be used to find potentially
predictive biomarkers
\citep{robins2008,tian2014,luedtke2016,chen2017,zhao2018,wager2018,fan2020,bahamyirou2022,hines2022}.

While CATE estimation procedures are demonstrably successful at predictive
biomarker discovery in settings where the number of features is small relative
to the sample size, it is not so in modern clinical trial in which the number
of features frequently exceeds the number of enrolled patients. The
high-dimensional nature of trial data make the CATE estimation problem
particularly difficult. Methods proposed for this setting must rely on
convenient --- and sometimes unverifiable --- assumptions about the underlying
data-generating process. Examples are sparsity, linear associations, and
negligible dependence structures
\citep{tian2014,chen2017,zhao2018,fan2020,bahamyirou2022}.
When these assumptions are violated, as is the case when, for example, the set
of biomarkers is comprised of gene expression data, the CATE estimate will be
biased but may be viable. The biomarkers designated as predictive, however,
will likely be false positives (as demonstrated in
Section~\ref{sec:simulations}).

\citet{hines2022} recently proposed a collection of variable 
importance parameters that assess the impact of variables, either
individually or in predefined sets, on the variance of the CATE. These
parameters are based on popular variable-dropout procedures and on
previous work about the variance of the conditional treatment effect
\citep{levy2021}. While the proposed estimators of these parameters are
consistent and asymptotically linear under non-restrictive assumptions
about the data-generating process, \citet{hines2022} note that
quantifying treatment effect modification in this way is misleading when
variables are highly correlated. Dropout-based importance metrics may
also be deceiving when there are many variables; other features may act
as surrogates for the omitted covariate(s) \citep[Chap. 15]{esl2009}.
This framework is therefore inappropriate for the discovery of predictive
biomarkers in high dimensions.

Still other procedures not relying on CATE estimation have recently been
proposed. \citet{sechidis2018} developed an information-theoretic
approach for identifying these treatment effect modifiers, though the
statistical properties of the procedure are not established and the
simulations do not consider high-dimensional data. \citet{zhu2022}
recently developed a penalized linear modelling method for the
identification of predictive biomarkers in high dimensions that accounts
for the biomarker correlation structure. Like the previous method,
however, no formal statistical guarantees are provided.

Myriad methods attempting to identify high-dimensional interactions more
generally might also be considered for our task \citep[for
example,][]{hao2014, jiang2014, tang2020}. They too generally rely on
untenable simplifying assumptions about the data-generating process. These
include, but are not limited to, assumptions of normality, sparsity of the
main effects, sparsity of interaction effects, and bounds on the condition
number of the biomarker covariance matrix. Large sets of biomarkers are again
unlikely to satisfy such conditions, barring these methods' use for
predictive biomarker discovery in high-dimensions.

A simpler alternative is to fit individual (generalized) linear models of
the outcome for each biomarker. Each model is comprised of the
biomarker's main effect and a treatment-biomarker interaction term. The
effect size estimate of the latter serves as a measure importance; larger
magnitudes equate to increased treatment effect modification. Hypothesis
testing about these treatment-biomarker interaction effects is also
possible. As with CATE estimation methods, however, this simple approach
imposes stringent parametric conditions on the data-generating process.
When the outcome is continuous, for example, inference is only possible
when all marginal biomarker-outcome relationships are truly linear. In
small samples, an additional assumption of Gaussian error terms is needed
for valid hypothesis testing. Violation of these unrealistic conditions
again produces unreliable predictive biomarker identification.

A lack of Type-I error control has marked repercussions in many biomedical
applications. In drug target discovery, limited resources are wasted by
performing biological follow-up experiments on false positives. In diagnostic
development, the inclusion of non-predictive biomarkers may dilute the signal
from truly informative ones. In a sequencing-based diagnostic, invalid
biomarkers will compete with others for sequencing reads, reducing the
sequencing depth and, thereby, the quantification accuracy of predictive
biomarkers. These failings have direct, detrimental effects on patient health
outcomes.

Motivated by these drawbacks, we present in this work a flexible approach for
directly assessing the predictive potential of individual biomarkers. That is,
we estimate (a transformation of) each biomarker's \textit{univariate} CATE, a
novel variable importance parameter for treatment effect modification. What is
more, our procedure permits the formal statistical testing of these biomarkers'
predictive effects under non-restrictive assumptions about the underlying
data-generating process, and we find that it controls the false discovery rate
(FDR) at the nominal level for realistic sample sizes. We also demonstrate on
real-world data that our method provides reasonable sub-population
identification results when combined with standard clustering approaches.

We emphasize that our framework is not a competitor of treatment rule
estimation procedures, it is complementary. The estimation of the CATE and the
identification of predictive biomarkers are related but distinct pursuits. To
highlight this, we might consider a two-step procedure wherein the full set of
biomarkers is filtered using our method, and then the CATE is estimated using
the remaining features. The benefits of such a strategy are numerous. The
results of the initial stage can help assess whether the assumption of sparsity
used by existing methods is tenable, and therefore whether estimating the CATE
is feasible. If not, then the ranking of biomarkers might still provide
biological or clinical insight, or motivate further study. If so, the CATE may
be estimated more accurately, thanks to the reduced number of features
considered, using flexible methods like those of \citet{tian2014},
\citet{luedtke2016}, or \citet{wager2018}. Further, the rankings generated in
the initial stage can impart intuition about the otherwise uninterpretable
treatment rule produced by ``black-box'' methods.

The remainder of the manuscript is organized as follows: In
Section~\ref{sec:prob-form}, the estimation setting and problem are detailed in
statistical terms. Section~\ref{sec:est-inf} then describes the proposed
inferential procedures. The asymptotic behavior of our method is then verified
empirically through a comprehensive simulation study in
Section~\ref{sec:simulations}. Application of the proposed approach to clinical
trial data then follows in Section~\ref{sec:app-rct}. We end with a brief
discussion of the method in Section~\ref{sec:discussion}. Throughout, we
emphasize inference about the univariate CATEs in a randomized control trial
setting, though some remarks on its application to observational data are also
provided.

\section{Variable Importance Parameters}\label{sec:prob-form}

Consider $n$ identically and independently distributed (\textit{i.i.d.}) random
vectors $X_i = (W_i, A_i, Y^{(1)}_i, Y^{(0)}_i) \sim P_X$, $i = 1, \ldots, n$,
corresponding to complete but unobserved data generated by participants in an
idealized randomized control trial or observational study. We drop the indices
for notational convenience where possible throughout the remainder of the
article. Here, $W = (V, B)$ is a $(q+p)$-length random vector of $q$
pre-treatment covariates, $V$, like location and income, and $p$ pre-treatment
biomarkers, $B$, such as gene expression data, $A$ is a binary random variable
representing a treatment assignment, and $Y^{(1)}$ and $Y^{(0)}$ are random
variables corresponding to the potential outcomes of clinical interest under
both treatment and control conditions, respectively \citep{rubin1974}. The
number of biomarkers $p$ is assumed to be approximately equal to or larger than
$n$. Generally, only one potential outcome is observed per unit. We ignore this
point for now, and return to it in the next section.

Clinically relevant predictive biomarkers are often those that have a strong
influence on the outcome of interest on the absolute scale. As such, an
ideal target of inference when these outcomes are continuous and the number of
covariates small is the CATE conditioning on the set of biomarkers:
\begin{equation*}
  \mathbb{E}_{P_X}\left[
    Y^{(1)} - Y^{(0)} \big| B = b
  \right].
\end{equation*}
For reasons previously discussed, however, accurate and interpretable
estimation of this parameter is generally challenging when $p$ is large,
preventing the accurate recovery of predictive biomarkers.

Indexing the biomarkers of by $j = 1, \ldots, p$, such that $B = (B_{1},
\ldots, B_{p})$, centering them such that $\mathbb{E}_{P_X}[B_{j}] = 0$, and
assuming that $\mathbb{E}_{P_X}[B^2_{j}] > 0$, we instead target the full-data
variable importance parameter $\Psi^F(P_X) = (\Psi^F_1(P_X), \ldots,
\Psi_p^F(P_X))$ where
\begin{equation}\label{eq:uni-cate}
  \Psi_{j}^F(P_X) \equiv \frac{\mathbb{E}_{P_X}\left[\left(Y^{(1)} -
    Y^{(0)}\right)B_{j}\right]} {\mathbb{E}_{P_X}\left[B_{j}^2\right]}.
\end{equation}
Under the assumption that the mean difference in potential outcomes admits a
linear form when conditioning on any given $B_j$, $\Psi^F(P_X)$ is the vector
of expected simple linear regression coefficients produced by regressing the
difference in potential outcomes against each biomarker. While the true
relationship between the difference of potential outcomes and a predictive
biomarker is almost surely nonlinear, $\Psi^F(P_X)$ is a generally informative
target of inference. Biomarkers with the largest absolute values in
$\Psi^F(P_X)$ generally modify the effect of treatment the most.

Analogous simplifications of the high-dimensional regression problem are
applicable to other types of outcome variables. For binary outcomes, we might
similarly wish to quantify the importance of biomarkers on the absolute
risk scale using a slightly modified univariate CATE parameter,
$\Psi^{F\text{(binary)}}(P_X) = (\Psi_1^{F\text{(binary)}}(P_X), \ldots,
\Psi_p^{F\text{(binary)}}(P_X))$, where:
\begin{equation}\label{eq:uni-bin-cate}
  \Psi_{j}^{F\text{(binary)}}(P_X) \equiv \frac{\mathbb{E}_{P_x}\left[
    \left(\mathbb{P}_{P_x}\left[Y^{(1)} = 1 \big| W\right] -
    \mathbb{P}_{P_X}\left[Y^{(0)} = 1 \big| W\right]\right)B_{j}\right]}
    {\mathbb{E}_{P_X}\left[B_{j}^2\right]}
\end{equation}
for centered biomarkers $j = 1, \ldots, p$. This is, in fact, the same
parameter as $\Psi^F(P_X)$ but presented in a more intuitive form for the
binary outcome context: assuming a linear relationship between the difference
of the potential outcomes' probability of success and the covariates, this
variable importance parameter consists of the simple linear regression
coefficients of the difference in the conditional potential outcome success
probabilities regressed on each biomarker. Again, the true relationship between
the difference of potential outcome probabilities and covariates is unlikely to
be linear. Nevertheless, this parameter is telling of biomarkers' predictive
capacities. 

We stress that the parameters in Equations~\eqref{eq:uni-cate} and
\eqref{eq:uni-bin-cate} are reasonable approximations of all but pathological
treatment effect modification relationships; they summarize the true,
marginal functional parameters using interpretable linear models. A case in
which $\Psi^F_j(P_X)$ will fail to capture treatment effect modification due
to biomarker $j$ is when the $\mathbb{E}_{P_0}[Y^{(1)}-Y^{(0)}|B_j]$ is
parabolic: the orthogonal projection of $Y^{(1)}-Y^{(0)}$ onto $B_j$ produces
a variable importance parameter value of zero. If such relationships are
suspected, however, it suffices to target the corresponding variable
importance parameters of the squared biomarkers. Analogous parameters based
on transformations of the biomarkers should be considered when the
data-generating process is assumed to possess other similarly troublesome
nonlinearities.

\section{Inference}\label{sec:est-inf}

As previously mentioned, only one of the potential outcomes, $Y^{(0)}$ or
$Y^{(1)}$, is observed per unit. Instead of $\{X_i\}_{i=1}^n$, we have access
to $n$ i.i.d. random observations $O = (W, A, Y) \sim P_0 \in \mathcal{M}$,
where $W$ and and $A$ are defined as before, and $Y = AY^{(1)} + (1-A)Y^{(0)}$
is a continuous or binary random outcome variable. $P_0$ is the unknown
data-generating distribution of the observed data that is fully determined by
$P_X$ and the (conditional) treatment assignment distribution $g_{A|W}$. That
is, $P_0$ is an element of the nonparametric statistical model
$\mathcal{M}=\{P_{P_X, g_{A|W}}: P_X \in \mathcal{M}_X, g_{A|W}\}$. In a
perfect RCT, $g_{A|W} = g_A = \text{Bernoulli}(0.5)$. The challenge therefore
lies in estimating the full-data, causal parameter of
Equations~\eqref{eq:uni-cate} and \eqref{eq:uni-bin-cate} with the observed
data; it is generally impossible without making additional assumptions about
$P_0$. We begin by providing such identification conditions.

Throughout the remainder of the text, we represent the empirical distribution
of $P_0$ by $P_n$, the conditional outcome regression function by $\bar{Q}_0(a,
w) \equiv \mathbb{E}_{P_0}[Y|A = a, W = w]$, and the treatment assignment
mechanism by $g_0(a, w) \equiv \mathbb{P}_{P_0}[A = a | W = w]$. Where
possible, we simplify notation further by writing $\bar{Q}_0(a, w)$ and $g_0(a,
w)$ as $\bar{Q}_0$ and $g_0$, respectively. All proofs are provided in
Section~\ref{sec:proofs} of the Appendix.

\begin{assumption}\label{ass:unmeasured-confounding}
  No unmeasured confounding: $Y^{(a)} \perp A | W$ for $a = \{0, 1\}$.
\end{assumption}
\begin{assumption}\label{ass:positivity}
  Positivity: There exists some constant $\epsilon > 0$ such that
  $\mathbb{P}_{P_0}[\epsilon < g_0(1, W) < 1-\epsilon] = 1$.
\end{assumption}

A\ref{ass:unmeasured-confounding} assures that there are no
unmeasured confounders of treatment and outcome, allowing for treatment
allocation to be viewed as the product of a randomized experiment.
A\ref{ass:positivity} is an overlapping support condition stating
that all observations may be assigned to either treatment condition regardless
of covariates. These conditions, regularly cited in the causal inference
literature, are generally satisfied in randomized control trials. Altogether,
they lead to the following result:
\begin{theorem}\label{thm:identification}
  Under the conditions of A\ref{ass:unmeasured-confounding} and
  A\ref{ass:positivity}, letting $\mathbb{E}_{P_0}[B_{j}] = 0$, and assuming
  that $\mathbb{E}_{P_0}[B^2_{j}] > 0$,
  \begin{equation}\label{eq:cont-out-estimand}
    \begin{split}
      \Psi_j(P_0)
      & \equiv \frac{\mathbb{E}_{P_0}\left[\left(\bar{Q}_0(1, W) -
        \bar{Q}_0(0, W)\right)B_{j}\right]}
        {\mathbb{E}_{P_0}\left[B_{j}^2\right]} \\
      & = \Psi_j^F(P_X)
    \end{split}
  \end{equation}
  for $j = 1, \ldots, p$ such that $\Psi(P_0) = (\Psi_1(P_0), \ldots,
  \Psi_p(P_0)$. Further, define the Augmented Inverse Probability Weight (AIPW)
  transform as
  \begin{equation}\label{eq:aipw-transform}
    T_a(O; \bar{Q}_0, g_0) =
    \frac{I(A=a)}{g_0(A, W)}
    (Y - \bar{Q}_0(A, W)) + \bar{Q}_0(a, W),
  \end{equation}
  and let $\tilde{T}(O; P_0) = T_1(O; \bar{Q}_0, g_0) - T_0(O; \bar{Q}_0,
  g_0)$. Then, the efficient influence function (EIF) of $\Psi_j(P)$ for $P \in
  \mathcal{M}$ and $j=1, \ldots, p$ is given by
  \begin{equation}\label{eq:cont-out-eif}
    \text{EIF}_j(O; P) \equiv \frac{
      \left(\tilde{T}(O; P) - \Psi_j(P)B_{j}\right)B_{j}}
      {\mathbb{E}_{P}\left[B_{j}^2\right]}.
  \end{equation}
\end{theorem}

Having established the conditions under which $\Psi^F(P_X)$ can be estimated
from the observed data, we now focus on inference about $\Psi(P_0)$. The EIF of
Equation~\eqref{eq:cont-out-eif} informs the construction of nonparametric
efficient estimators of $\Psi_j(P_0)$ under non-restrictive assumptions about
the data-generating process \citep{bickel1993,hines2021}. Many approaches exist
for deriving these efficient estimators, such as one-step estimation
\citep{pfanzagl1985, bickel1993}, estimating equations
\citep{laan2003,cherno2017,cherno2018}, or targeted maximum likelihood
estimation \citep{laan2006,laan2011,laan2018}. We use the, in this case,
straightforward method of estimating equations. The resulting estimator is
intuitive: it corresponds to the estimator of the simple linear regression
coefficient of centered biomarker $j$ regressed on the adjusted predicted
differences in potential outcomes. Further, it is identical to the one-step
estimator.

\begin{corollary}\label{cor:estimator}
  Let $P_m$ be the empirical distribution of another dataset of $m$ random
  observations distributed according to $P_0$ and distinct of $P_n$. If such a
  dataset is not available, it might be generated using sample-splitting
  techniques. We require that the size of $P_m$ grows linearly with the size of
  $P_n$. That is, $O(m) = O(n)$. This is trivially accomplished when using
  most sample-splitting frameworks, like K-fold cross-validation. Then define
  $\bar{Q}_m$ and $g_m$ as estimates of the nuisance parameters $\bar{Q}_0$ and
  $g_0$ fit to $P_m$. The estimating equation estimator of $\Psi_j(P_0)$ is
  then given by:
  \begin{equation}\label{eq:cont-out-ee-est}
    \Psi^{(ee)}_j(P_n; P_m) = \frac{\sum_{i=1}^n \tilde{T}(O_i;
    P_m)B_{ij}}{\sum_{i=1}^n B_{ij}^2},
  \end{equation}
  where we again assume that the biomarkers are centered such that
  $\sum B_{ij} = 0$. This estimator is double robust.
\end{corollary}

The double-robustness property signifies that $\Psi^{(\text{ee})}_j(P_n; P_m)$
is a consistent estimator of $\Psi_j(P_0)$ so long as either the estimator of
the conditional expectation or the estimator of the propensity score are
consistent. In particular, when $g_0$ is known, as in most clinical trials, it
is guaranteed to be consistent.

Under the following conditions, we can detail this estimator's limiting
distribution.
\begin{assumption}\label{ass:known-treatment}
    Known treatment assignment mechanism: $g_0$ is known.
\end{assumption}
\begin{assumption}\label{ass:double-rate-robustness}
  Nuisance parameter estimator convergence:
  $\lVert \bar{Q}_m - \bar{Q}_0 \rVert_2 \;
  \lVert g_m - g_0 \rVert_2 = o_{P}(n^{-1/2})$, where $\lVert \cdot \rVert_2$
  denotes the $L_2(P_0)$ norm.
\end{assumption}

\begin{theorem}\label{thm:inference}
  If A\ref{ass:known-treatment} or A\ref{ass:double-rate-robustness} are
  satisfied and $\mathbb{E}_{P_0}[B_j^2] > 0$ for $j=1,\ldots,p$, then
  \begin{equation}\label{eq:limiting-dist}
    \sqrt{n}\left(\Psi^{(ee)}_j(P_n; P_m) - \Psi_j(P_0)\right)
    \overset{D}{\rightarrow} N\left(0,
    \mathbb{V}_{P_0}\left[\text{EIF}_j(O; P_0)\right]\right).
  \end{equation}
\end{theorem}

Again, A\ref{ass:known-treatment} is generally satisfied in clinical
trials, implying that the estimating equation estimator of
Equation~\eqref{eq:cont-out-ee-est} is asymptotically linear. Valid hypothesis
testing is possible even when the conditional outcome regression is biased.
This results from the form of the EIF, and is discussed in the proof
(Section~\ref{sec:proofs} of the Appendix).

In observational settings, A\ref{ass:double-rate-robustness} requires
that the conditional outcome regression estimates and the treatment assignment
rule estimates converge in probability to their respective true parameters at a
rate faster than $n^{-1/4}$. When the number of biomarkers and covariates is
moderate relative to sample size, these conditions are typically satisfied by
estimating these parameters using flexible machine learning algorithms
\citep{laan2011} like the Super Learner of \citet{laan2007}. Relying on the
general asymptotic theory of cross-validated loss-based estimation
\citep{laan-dudoit2003}, the Super Learner method constructs a convex
combination of estimators from a pre-specified library that minimizes the
cross-validated risk of a pre-defined loss function. Even in a high-dimensional
setting where the number of biomarkers is far larger than $n$, recent results
about Random Forests \citep{wager2018} and deep neural networks
\citep{farrell2021} suggest conditions for which
A\ref{ass:double-rate-robustness} is satisfied. Generally, fast
convergence of these estimators in high dimensions requires strong smoothness
and sparsity assumptions about the underlying parameters \citep{hines2021}.  

Under A\ref{ass:unmeasured-confounding}, A\ref{ass:positivity}, and either
A\ref{ass:known-treatment} or A\ref{ass:double-rate-robustness},
Theorem~\ref{thm:inference} delivers the means by which to construct
$\alpha$-level Wald-type confidence intervals for $\Psi_j(P_0)$. However, the
estimator of Equation~\eqref{eq:cont-out-ee-est} and any accompanying testing
procedure require that the nuisance parameters be estimated on a separate
dataset.

Since practitioners rarely have access to two datasets from the same
data-generating process, we propose a cross-validated estimator that uses all
available data. Begin by randomly partitioning the $n$ observations of $P_n$
into $K$ independent validation sets $P_{n, 1}^1, \ldots, P_{n, K}^1$ of
approximately equal size. For $k=1,\ldots, K$, define the training set as, in a
slight abuse of notation, $P_{n, k}^0 = P_n \setminus P_{n,k}^1$. Then the
cross-validated estimator of $\Psi_j(P_0)$ is defined as:
\begin{equation}\label{eq:cv-estimator}
  \Psi_j^{(\text{CV})}(P_n) = \frac{1}{K} \sum_{k=1}^K
  \frac{\sum_{i=1}^n I(O_i \in P_{n, k}^1)\tilde{T}(O_i; P_{n,k}^0)B_{ij}}
  {\sum_{i=1}^n I(O_i \in P_{n, k}^1) B_{ij}^2},
\end{equation}
and has the same limiting distribution as $\Psi_j^\text{(ee)}(P_n;P_m)$ under
conditions consistent with those of either A\ref{ass:known-treatment} or
A\ref{ass:known-treatment}. The accompanying cross-validated estimator of the
EIF's standard deviation for biomarker $j$ is given by
\begin{equation*}
  \sigma^{\text{(CV)}}_j(P_n) = \left(\frac{1}{K} \sum_{k=1}^K\left[
  \frac{1}{\sum_{i=1}^n I(O_i \in P_{n, k}^1)}
    \sum_{i=1}^n I(O_i \in P_{n, k}^1)
    \left(\text{EIF}(O_i; P_{n,k}^0)\right)^2\right]\right)^{1/2}.
\end{equation*}
The $\alpha$-level Wald-type confidence intervals for $\Psi_j(P_0)$ are then
constructed as
\begin{equation*}
  \Psi^{\text{(CV)}}_j(P_n) \pm
  \frac{z_{(1-\alpha/2)} \sigma_{_j}^{(\text{CV})}(P_n)}{\sqrt{n}},
\end{equation*}
where $z_{(1-\alpha/2)}$ is the $(1-\alpha/2)^\text{th}$ quantile of a standard
Normal distribution. Inference about $\Psi_j(P_0)$ is therefore made possible
under non-restrictive assumptions about the data-generating process when using
data-adaptive methods and cross-validation to the estimate nuisance parameters.
Even in small samples where the limiting properties of
$\Psi_j^\text{(ee)}(P_n;P_m)$ might not be attained, the generalized variance
moderation technique of \citet{hejazi2022} can be used for Type-I error
control.

In summary, we take as target of inference the simple linear regression slopes
of the difference of predicted outcomes under treatment and control conditions
regressed on each biomarker. We suggest that these parameters be estimated by
learning the conditional outcome regression and treatment assignment rule (if
necessary), using them to predict the potential outcomes, and then fitting
simple linear regressions to the difference in predicted potential outcomes as
a function of each centered biomarker. Under the conditions outlined in
A\ref{ass:unmeasured-confounding}, A\ref{ass:positivity}, and
A\ref{ass:known-treatment} or A\ref{ass:double-rate-robustness}, we prove that
our estimator targets the causal parameter of interest and is asymptotically
linear, providing a straightforward statistical test to assess whether a
biomarker modifies the treatment effect. Even when the causal inference
conditions of A\ref{ass:unmeasured-confounding} and A\ref{ass:positivity} are
not satisfied, $\Psi(P_0)$ remains an interpretable statistical
parameter. It captures the strength of treatment-biomarker interactions in high
dimensions, and inference about it can be performed using the same
cross-validated procedure.

\section{Simulation Study}\label{sec:simulations}

\subsection{Details}

This work is motivated by the need to identify predictive biomarkers in
clinical trials. Of particular interest is their detection for drug target
discovery and diagnostic assay development. The former requires the
identification of biomarkers causally related to the outcome of interest,
whereas the latter seeks a small set of strongly predictive biomarkers. We
therefore focus on these applications throughout the simulation study.

A varied collection of data-generating processes, defined below, are considered
to demonstrate that the theoretical guarantees outlined in the previous section
are achieved for a range of functional forms of the conditional outcome
regression. Recall that $Y$ corresponds to the outcome, $A$ the treatment
assignment, $W$ the covariates, and $B$ the biomarkers, a subset of the
covariates. The treatment assignment rules, $g_0$, are treated as known. 

\begin{itemize}

  \item Class 1: Moderate dimensions, non-sparse treatment-biomarker effects
    vector with independent biomarkers

    \begin{itemize}
      
      \item Linear conditional outcome regression:
        \begin{align*}
          W = B & \sim N(0, I_{100 \times 100}) \\
          A | W = A & \sim \text{Bernoulli}(1/2) \\
          Y | A, W & \sim N\left(W^\top
            \left(\beta + I(A = 1)\gamma^{(1)} + I(A = 0)\gamma^{(0)}\right)
          , 1/2\right).
        \end{align*}
        Here, $\beta = (\beta_1, \ldots, \beta_{100})^\top$ such that $\beta_1
        = \ldots = \beta_{20} = 2$ and $\beta_{21} = \ldots = \beta_{100} = 0$,
        and $\gamma^{(a)} = (\gamma_1^{(a)}, \ldots, \gamma_{100}^{(a)})^\top$
        where $\gamma_1^{(1)} = \ldots = \gamma_{50}^{(1)} = 5$,
        $\gamma_1^{(0)} = \ldots = \gamma_{50}^{(0)} = -5$ and
        $\gamma_{51}^{(1)} = \ldots = \gamma_{100}^{(1)} = 0$ for $a = \{0,
        1\}$.

      \item Kinked conditional outcome regression: $W$ and $A$ are
        distributed as above. The conditional outcome is defined as
        \begin{equation*}
          Y | A, W \sim N\left(
            W^\top \left(I(A = 1)\gamma + I(A = 0)\;\text{diag}(I(W >
            0))\;\gamma\right), 1/2
          \right),
        \end{equation*}
        where $\gamma = (\gamma_1, \ldots, \gamma_{100})$, $\gamma_{1} = \ldots
        = \gamma_{50} = 10$, $\gamma_{51} = \ldots = \gamma_{100} = 0$, and
        $\text{diag}(\cdot)$ is a diagonal matrix whose diagonal equals the
        input vector.

      \item Nonlinear conditional outcome regression: $W$ and $A$ are
        distributed as above. Then,
        \begin{equation*}
          Y | A, W \sim N\left(
            \text{exp}\left\{\lvert W^\top \beta \rvert\right\}
            + I(A = 1)W^\top\gamma, 1/2
          \right),
        \end{equation*}
        where $\beta_1 = \ldots = \beta_{20} = 1$ and $\beta_{21} = \ldots =
        \beta_{100} = 0$, and where $\gamma_1 = \ldots = \gamma_{50} = 5$ and
        $\gamma_{51} = \ldots = \gamma_{100} = 0$.

    \end{itemize}

  \item Class 2: High dimensions, sparse treatment-biomarker effects vector
    with correlated biomarkers

    \begin{itemize}
      \item Linear conditional outcome regression:
        \begin{align*}
          C & \sim \text{Bernoulli}(1/2) \\
          W | C = B | C & \sim N(-I(C=0) + I(C=1), \Sigma_{500 \times 500})  \\
          A|W = A & \sim \text{Bernoulli}(1/2) \\
          Y | A, W & \sim N\left(
             W^\top\left(\beta + I(A = 1)\gamma\right), 1/2\right)
        \end{align*}
        Here, $C$ is an unobserved subgroup indicator, $\beta = (2, 2, 2, 2, 2,
        0, \ldots, 0)$, and $\gamma = (5, 5, 5, 5, 0, \ldots, 0)$. The
        biomarker covariance matrix, $\Sigma$, is the estimated gene expression
        correlation matrix of the 500 most variable genes taken from the
        tumours of patients with metastatic or recurrent colorectal cancer
        \citep{watanabe2011}. These genes were first clustered using
        hierarchical clustering based on their Euclidean distance with complete
        linkage, and the correlation matrix was then estimated using the
        cross-validated estimation procedure of \citet{boileau2021} implemented
        in the \texttt{cvCovEst} \texttt{R} package \citep{cvCovEst,rcore},
        relying on the banding and tapering estimators of \citet{bickel2008}
        and \citet{cai2010}, respectively. The gene expression data has been
        made available by the Bioconductor \citep{huber2015} experiment package
        \texttt{curatedCRCdata} \citep{parsana2021}.

      \item Kinked conditional outcome regression: $C$, $W$ and $A$ are
        distributed as above. The conditional outcome distribution is as
        follows: 
        \begin{equation*}
          Y | A, W \sim N\left(
            W^\top \left(I(A = 1)\gamma + I(A = 0)\;\text{diag}(I(W >
            0))\;\gamma\right), 1/2
          \right),
        \end{equation*}
        where $\gamma = (10, 10, 10, 10, 0, \ldots, 0)$.

      \item Nonlinear conditional outcome regression: $C$, $W$ and $A$ are
        distributed as above. Then,
        \begin{equation*}
          Y | A, W \sim N\left(
            \text{exp}\left\{\lvert W^\top\beta\rvert\right\}
            + I(A = 1)W^\top\gamma, 1/2
          \right),
        \end{equation*}
        where $\beta = (1, 1, 1, 1, 1, 0, \ldots, 0)$ and $\gamma = (5, 5,
        5, 5, 0, \ldots, 0)$.

    \end{itemize}

\end{itemize}

The first class of data-generating processes reflects the scenario in which a
set of biomarkers known to be associated with the outcome, perhaps based on
prior clinical or biological investigations, are assessed for potential
treatment-biomarker interactions. Since they have been cherry-picked, a
reasonable assumption is that a non-negligible proportion of these biomarkers
modify the effect of treatment on the outcome of interest. The second set of
data-generating processes is representative of exploratory scenarios wherein a
vast number of biomarkers, like tumor gene expression data collected prior to
the start of treatment, are explored for strong effect modification. Further,
these data-generating processes contain two subgroups, representing, for
example, unknown patient subpopulations in a clinical trial. These models each
possess four non-zero treatment-biomarker interactions in the leading entries
of $\gamma$ or $\gamma^{(0)}$ and $\gamma^{(1)}$. The biomarkers that modify
the treatment effect are correlated mimicking a small gene set.

\begin{figure}
  \center
  \includegraphics[width=\textwidth]{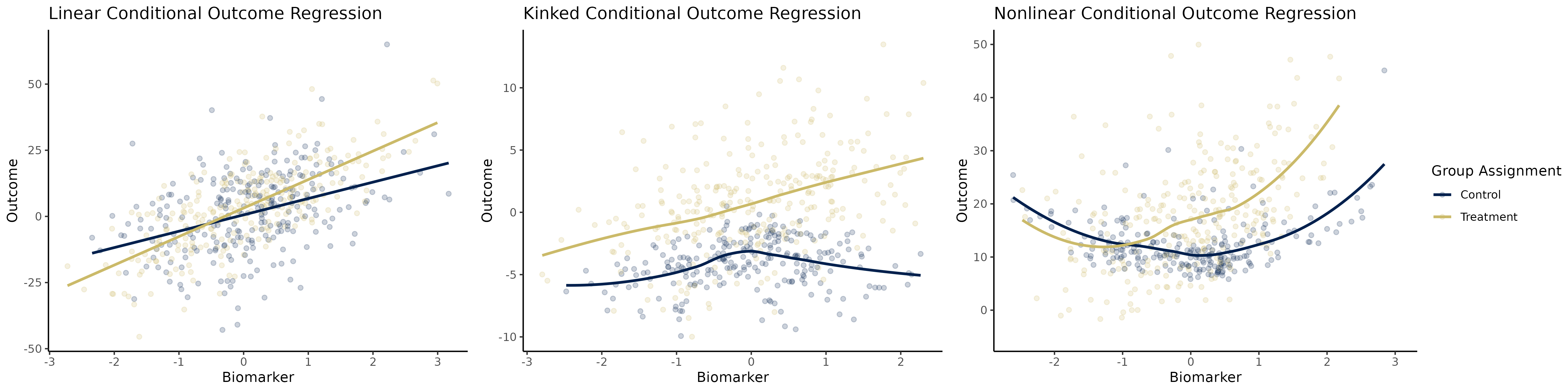}
  \caption{Sketches of predictive biomarkers' marginal relationships with the
    outcome variable for the considered conditional outcome regression models.}
  \label{fig:sketch-dgps}
\end{figure}
The collections of moderate and high-dimensional data-generating processes each
contain three outcome regression models. Their sketches are provided in
Figure~\ref{fig:sketch-dgps}. The simplest ``linear'' models correspond to the
functional form assumed by many existing high-dimensional CATE estimation
procedures for a continuous outcome
\citep{tian2014,chen2017,zhao2018,ning2020}. The ``kinked'' data-generating
processes are named so for the kink in the marginal conditional outcome
regression of its predictive biomarkers. These marginal relationships are
representative of predictive biomarkers in clinical trials assessing the
efficacy of the standard of care against combinations of the standard of care
and another drug, and where the treatment group outperforms the control group
in all biomarker defined subpopulations, but with different treatment effect
sizes. Finally, the ``nonlinear'' data-generating mechanisms represent those
whose conditional outcome regressions deviate most from assumptions of
linearity. We expect these to pose the greatest challenge with respect to
identifying predictive biomarkers. We note that the linear conditional outcome
regression models are not identifiable, but this is not a concern for
generative purposes.

Two hundred datasets of 125, 250, and 500 observations were generated for each
of these data-generating processes --- 3,600 in all --- by sampling without
replacement from simulated populations of 100,000 observations. Each model's
$\Psi(P_0)$ was computed from its respective population. These random samples
and estimands are used in the following subsections to assess the finite sample
performance of our proposed procedure and to benchmark its ability to discover
predictive biomarkers against that of popular CATE estimation methods.

The cross-validated estimator of Equation~\eqref{eq:cv-estimator} is used to
estimate the vector of univariate CATE simple linear regression coefficients in
the simulated datasets using 5-fold cross-validation. Throughout the remainder
of the text, we refer to our proposed method as \textit{uniCATE}.
\citet{laan2007}'s Super Learner procedure is used to estimate the conditional
outcome regressions. The library of candidate algorithms is made up of ordinary
linear, LASSO, and elastic net regressions \citep{tibshirani1996,zou2005},
polynomial splines \citep{stone1997}, XGboost \citep{chen2016}, Random Forests
\citep{breiman2001}, and the mean model.

\subsection{Bias and Variance of Univariate CATE Estimator}

\begin{figure}
\centering
  \includegraphics[width=0.7\textwidth]{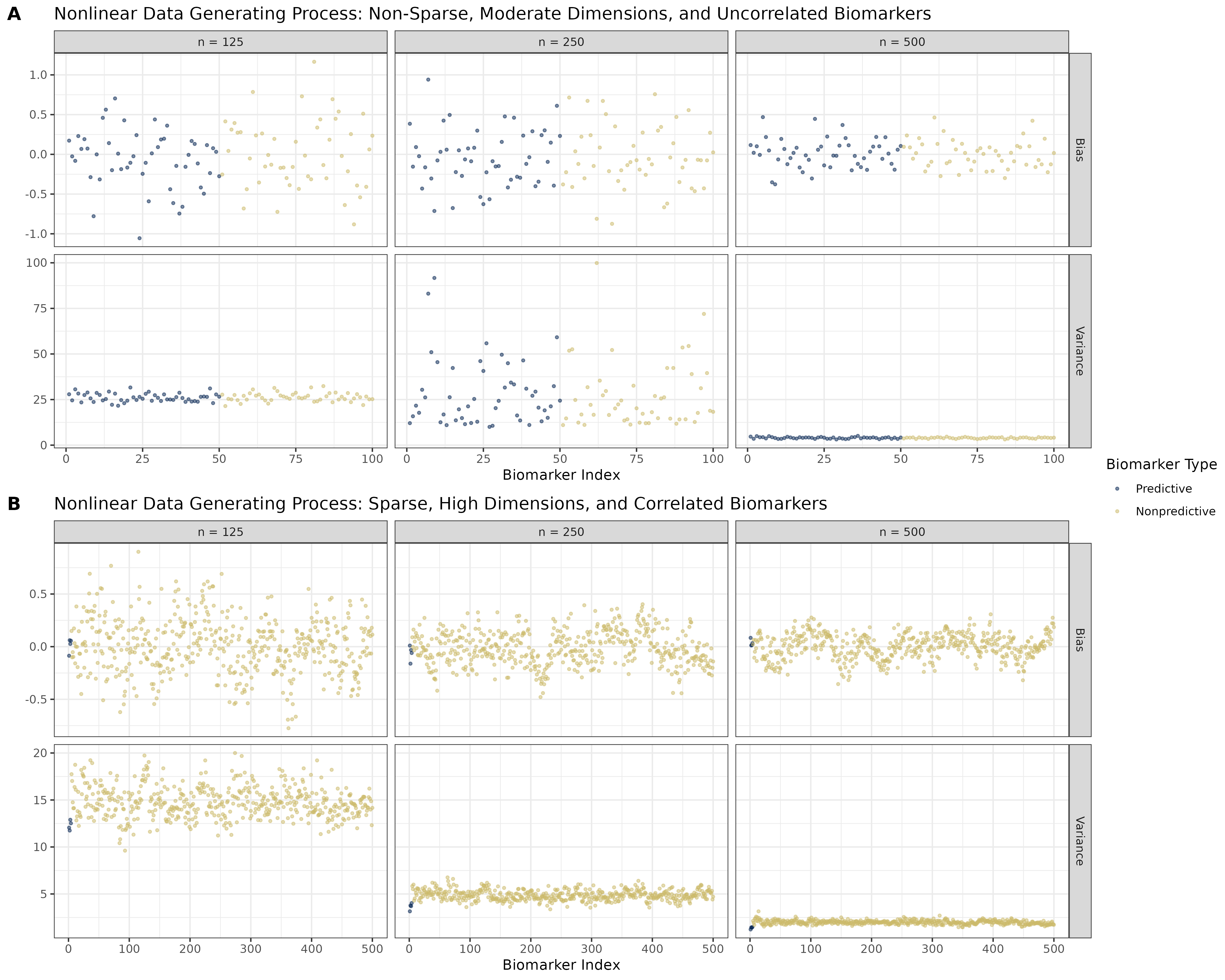}
  \caption{The empirical biases and variances of uniCATE estimates for all
  biomarkers across all simulation scenarios with a nonlinear conditional
  outcome regression. Biomarkers colored blue are truly predictive, and those
  colored gold are nonpredictive.}
  \label{fig:nonlinear-bias-variance}
\end{figure}

The theoretical results of Section~\ref{sec:est-inf} are asymptotic, yet many
clinical trials are made up of a small to moderate numbers of participants. We
therefore verify that uniCATE's estimates, metrics of biomarkers' predictive
importance, are accurate when computed under realistic sample sizes. We
computed the empirical bias and variance of the cross-validated estimator when
applied to each data-generating process and at each simulated sample size. The
results of our analysis of the nonlinear models are presented in
Figure~\ref{fig:nonlinear-bias-variance}. Those of the linear and kinked
models, presented in Figures~\ref{fig:linear-bias-variance}
and \ref{fig:kinked-bias-variance}, respectively, are virtually identical.

We find that uniCATE is approximately unbiased across sample sizes regardless
of the conditional outcome regressions' complexities, as suggested by
Theorem~\ref{thm:inference}. However, the estimator is highly variable in the
moderate dimension, non-sparse (e.g. Figure~\ref{fig:nonlinear-bias-variance}A)
scenarios when $n=125$ and $250$, and somewhat variable when $n=125$ in the
high dimension, sparse data-generating processes (e.g.
Figure~\ref{fig:nonlinear-bias-variance}B). As expected, the empirical variance
of the estimator decreases drastically in all simulation settings as sample
sizes increase.

This is encouraging for diagnostic biomarker assay development: the ranking of
predictive biomarkers reported by uniCATE is reliable under realistic sample
sizes and data-generating processes. These results suggest that our method
accurately and precisely evaluates biomarkers with respect to their predictive
abilities when the number of truly predictive biomarkers is small in samples
possessing as few as 250 observations. Similar behavior is observed when there
are a large number of predictive biomarkers in trials of 500 subjects or more.

\subsection{Type-I Error Control}

\begin{figure}
  \centering
  \includegraphics[width=0.8\textwidth]{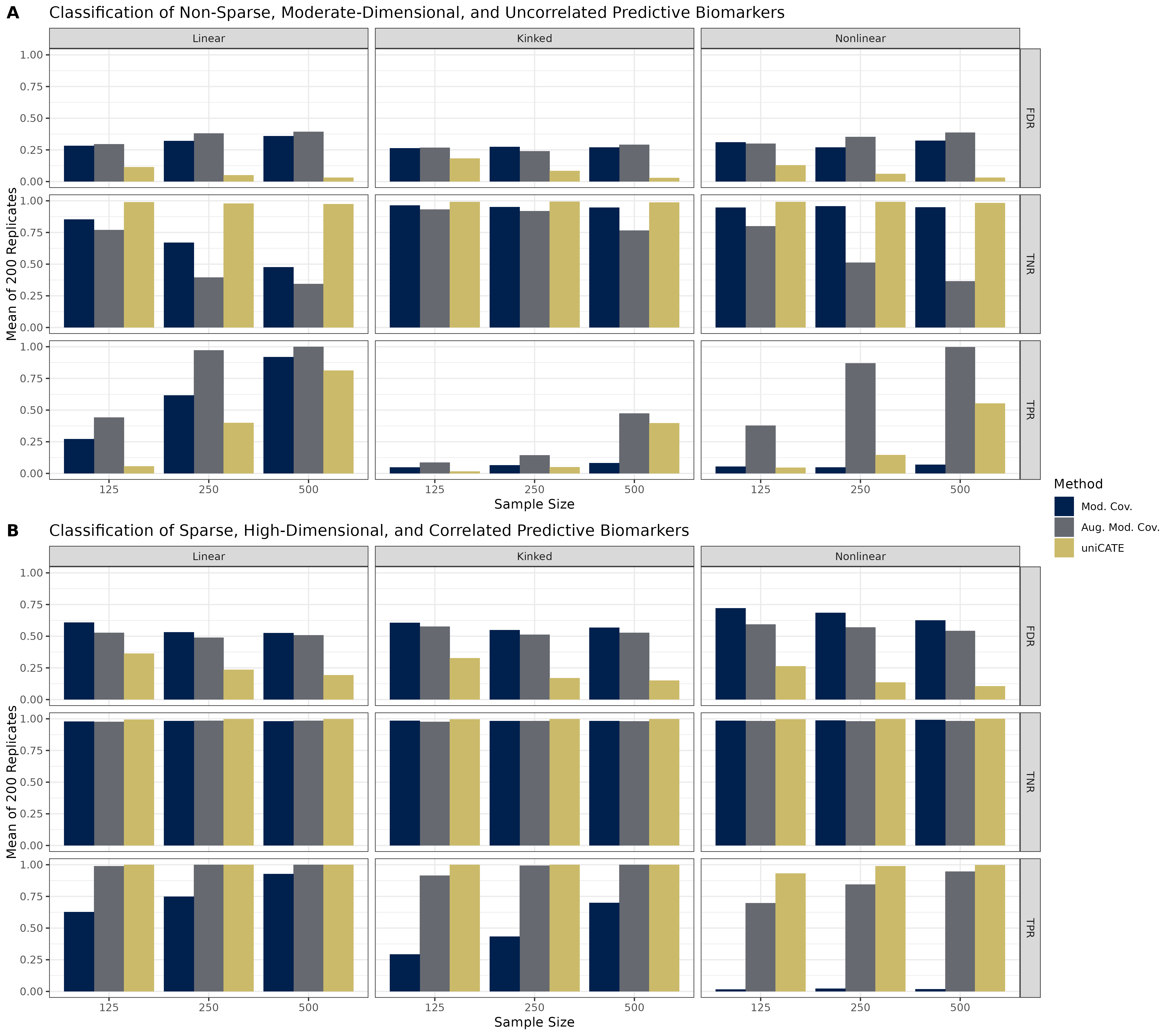}
  \caption{The empirical predictive biomarker classification results for the
  moderate dimensions, non-sparse treatment-biomarker interaction settings with
  uncorrelated biomarkers (A) and the high-dimension, sparse treatment-biomarker
  interaction settings with correlated biomarkers(B).}
  \label{fig:classification-results}
\end{figure}

In addition to evaluating the accuracy of uniCATE's estimates, we assess the
method's ability to distinguish predictive biomarkers from non-predictive
biomarkers. This is of particular importance in applications requiring the
reduction of the pool of potential predictive biomarkers, as in the development
of diagnostic assays, or generating hypotheses for biological and clinical
validation in drug target discovery. We therefore evaluate uniCATE's Type-I
error rate control across the simulation scenarios using a target FDR
\citep{benjamini1995} of 5\%. The inferential procedure described in
Section~\ref{sec:est-inf} is used to test whether predictive biomarkers' linear
approximations of the univariate CATE are significantly different from zero.
Nominal $p$-values are adjusted using the FDR-controlling procedure of
\citet{benjamini1995}. We note that nominal FDR control is not guaranteed by
this adjustment method in the high-dimensional simulations because of the
biomarkers' correlation structure. The results are presented in
Figure~\ref{fig:classification-results}.

Our method's capacity to identify predictive biomarkers was compared to that of
popular CATE estimation methods: the modified covariates approach and its
augmented counterpart \citep{tian2014,chen2017}. Briefly, the former directly
estimates the linear model coefficients of the treatment-biomarker
interactions, using a linear working model for these terms, without having to
model or estimate the main effects. While \citet{tian2014}'s method is flexible
since it avoids making any assumptions about the functional form of the main
biomarker effects, it can lack precision in small-sample, high-dimensional
settings. \citet{tian2014} and \citet{chen2017} therefore proposed
``augmented'' versions of this method that explicitly account for this source
of variation. While \citeauthor{tian2014}'s [\citeyear{tian2014}] and
\citeauthor{chen2017}'s [\citeyear{chen2017}] augmentation procedures differ,
they are identical in the randomized control trials with continuous outcome
variables \citep{chen2017}: they are equivalent to fitting a (penalized)
multivariate linear regression with treatment-biomarker interaction terms.

We again emphasize that these methods are not true competitors of our
procedure. Their primary goal, CATE estimation, differs from that of uniCATE.
However, \citet{tian2014} and \citet{chen2017} demonstrated that the
(augmented) modified covariates approach could identify potentially predictive
biomarkers when fit using regularized linear regressions like the LASSO.
Biomarkers with non-zero treatment-biomarker interaction coefficient estimates
are classified as predictive. We therefore applied these approaches, using
10-fold cross-validation to select the LASSO hyperparameters, to all simulated
datasets. The implementations of these estimators provided by the
\texttt{personalized} \texttt{R} software package \citep{huling2021} was used.

The results pertaining to the moderate dimension simulations ($p = 100$) with
$50$ predictive biomarkers are presented in
Figure~\ref{fig:classification-results}A. Only uniCATE is capable of
controlling the Type-I error rate; it approximately achieves the nominal FDR of
5\% in all settings with samples sizes of 250 and above. The modified
covariates approach and its augmented counterpart possess FDRs no lower than
25\% across all scenarios. Indeed, their control of Type-I error generally
worsens as sample size increases. Our method's superior performance with
respect to FDR control is likely due to its conservativeness: many of the
predictive biomarkers are not recognized in smaller sample-size settings. As
sample size grows, however, so too does its true positive rate (TPR) while
maintaining a near perfect true negative rate (TNR). When $n = 500$, uniCATE
generally identifies close to or more predictive biomarkers than the modified
covariates approach, and nearly as many as the augmented modified covariates
method.

Our procedure's performance with respect to FDR control is again superior to
that of the CATE estimation approaches in the high dimensional simulation
scenarios with $500$ biomarkers (Figure~\ref{fig:classification-results}B).
While the adjustment procedure of \citet{benjamini1995} does not guarantee FDR
control at the desired rate in these scenarios due to the correlation structure
of the tests, it is nearly achieved in larger sample sizes. uniCATE also
marginally outperforms other approaches in terms of the TNR. Unlike in the
moderate sample-size simulations, however, our method identifies predictive
biomarkers more efficiently than the other procedures considered.

These results demonstrate that uniCATE recovers truly predictive biomarkers
more reliably than interpretable treatment rule estimators. In most simulation
scenarios, uniCATE provides well controlled Type-I error rates while its TNR
and TPR are comparable or superior to other methods. However, when the number
of truly predictive biomarkers is large and the sample size small, uniCATE's
biomarker classification will be conservative. In this setting, our method
still provides good Type-I error control, limiting the waste of resources on
the investigation of false positives, as would be the result if using existing
methods. If the investigator prefers a less conservative approach, since, for
example, the cost of follow-up experiments is low, the (augmented) modified
covariate approach may be considered instead.

\section{Application to IMmotion Trials}
\label{sec:app-rct}

Until recently, Tyrosine kinase inhibitors targeting vascular endothelial
growth factor (VEGF) were the standard of care for patients with metastatic
renal cell carcinoma (mRCC) \citep{rini2019}. Unfortunately, many patients with
mRCC find these treatments, like sunitinib, ineffective, and most develop a
resistance over time \citep{rini2009}. Immune checkpoint inhibitors like
atezolizumab can produce more durable results and improve overall survival in
pre-treated patients with mRCC \citep{motzer2015a,motzer2015b,mcdermott2018}.
A combination of atezolizumab and bevacizumab, the latter of which also binds
to VEGF, was shown to improve the objective response rate (ORR) in a Phase 1b
study \citep{wallin2016}. Objective response is a binary indicator of
clinically meaningful response to treatment. These findings were supported by a
Phase 2 study, IMmotion 150, which compared atezolizumab alone and in
combination with bevacizumab against sunitinib \citep{mcdermott2018}. In a
subsequent Phase 3 study, IMmotion 151 \citep{rini2019}, the atezolizumab and
bevacizumab combination improved progression free survival and objective
response over sunitinib in patients whose cancer cells expressed the programmed
death-1 ligand 1 (PD-L1), but not all of these patients showed benefit. These
results motivate the search for biomarkers that are more predictive of this
treatment's clinical benefit than PD-L1 expression.

Potentially predictive biomarkers were found by applying uniCATE to subsets of
the sunitinib ($n=71$) and atezolizumab-bevacizumab ($n=77$) treatment arms of
the IMmotion 150 trial. Only patients with pre-treatment tumor RNA-seq samples
were included. The 500 most variable genes based on this log-transformed
RNA-seq data comprised the collection of potentially predictive biomarkers.
Details of the gene expression data collection and preparation have previously
been described by \citet{mcdermott2018}. Objective response was used as the
outcome variable. The conditional outcome regression model was fit with a
Super Learner whose library contained (penalized) GLMs with treatment-biomarker
interaction terms, XGboost models, Random Forest models, and the mean model. A
nominal FDR cutoff of 5\% was employed. The modified covariates and augmented
modified covariates approach for binary outcomes of \citet{tian2014} were also
applied to these data.

The uniCATE method identified 92 genes as predictive biomarkers, whereas the
modified covariates approach and its augmented counterpart identified 20 and 6,
respectively. All results are listed in Table~\ref{tab:predictive-biomarkers}.
That the approaches of \citet{tian2014} are more conservative than ours is a
reversal of Section~\ref{sec:simulations}'s simulation results, but may be
explained by the more complex correlation structure of these data. Indeed, the
former rely on sparse linear models which are known to select but a few
features from any given highly correlated set. Our procedure, however, uncovers
sets of correlated predictive biomarkers since their individual hypothesis
testing results will also be associated. This property is desirable when
analyzing genomic data as large gene sets permit more thorough biological
exploration and improved interpretation than do single, uncorrelated genes
\citep{subramanian2005}. The reporting of gene sets also improves the
reproducibility of findings \citep{subramanian2005}.

We performed a gene set enrichment analysis (GSEA) of gene ontology (GO) terms
with uniCATE's 92 predictive biomarkers using MSigDB
\citep{subramanian2005,tamayo2011}. The top results are presented in
Table~\ref{tab:GOBP}. We found that these genes are generally associated with
immune responses, including those mediated by B cells and lymphocytes. Similar
findings have been reported by \citet{au2021} in the context of clear cell
renal cell carcinoma patients' therapeutic responses to nivolumab, another
immune checkpoint inhibitor.

\begin{figure}
  \center
  \includegraphics[width=0.9\textwidth]{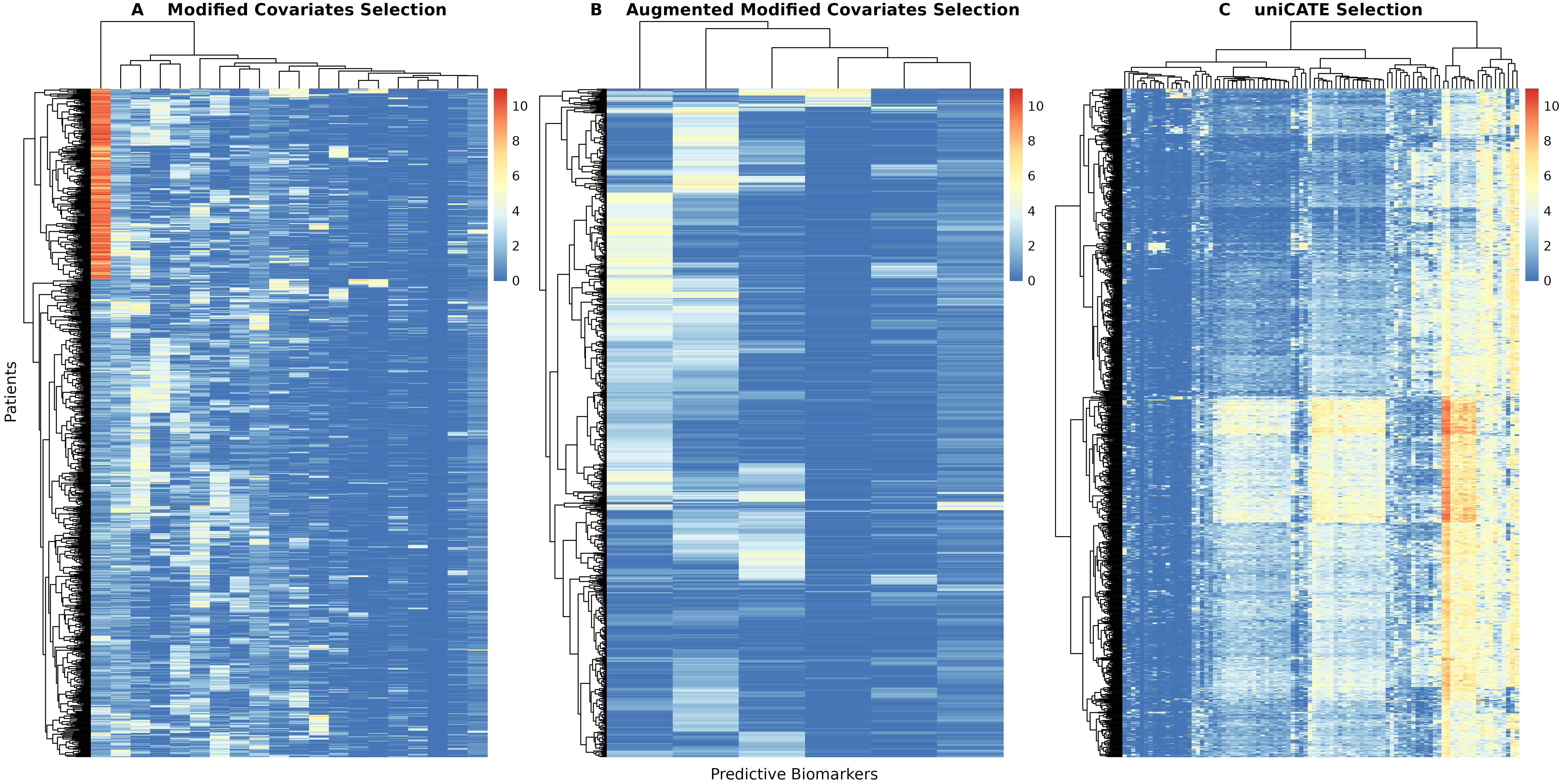}
  \caption{Heatmaps of the modified covariates approach's (A), augmented
  modified covariates approach's (B), and uniCATE procedure's (C) predictive
  biomarkers' log-transformed gene expression data from the IMmotion151 trial.
  Rows and columns are ordered via hierarchical clustering with complete
  linkage and Euclidean distance.}
  \label{fig:heatmaps}
\end{figure}

Now, having learned of potentially predictive biomarkers, we assessed how well
they delineate patient sub-populations in the IMmotion 151 study. This study's
subjects are believed to be drawn from the same population as those enrolled in
IMmotion 150. Eight hundred and ten subjects possessed baseline tumor gene
expression data for the 500 genes considered in our IMmotion 150 analysis: 406
in the atezolizumab-bevacizumab combination arm, and 404 in the sunitinib arm.
Figure~\ref{fig:heatmaps} presents the heatmaps of log-transformed gene
expression data for each methods' set of predictive genes. Subgroups are easily
discerned in uniCATE's heatmap, but not so much in the other procedures'. This
further emphasizes the benefits of uniCATE's capacity to identify sets of
correlated predictive biomarkers. Note that the most prominent cluster of
patients in the modified covariates method's heatmap is driven by the
\textit{XIST} gene. It is not selected by the augmented modified covariates
procedure or uniCATE. Upon further inspection, it does not appear to have a
strong predictive effect (Figure~\ref{fig:XIST}).

\begin{figure}
  \center
  \includegraphics[width=.7\textwidth]{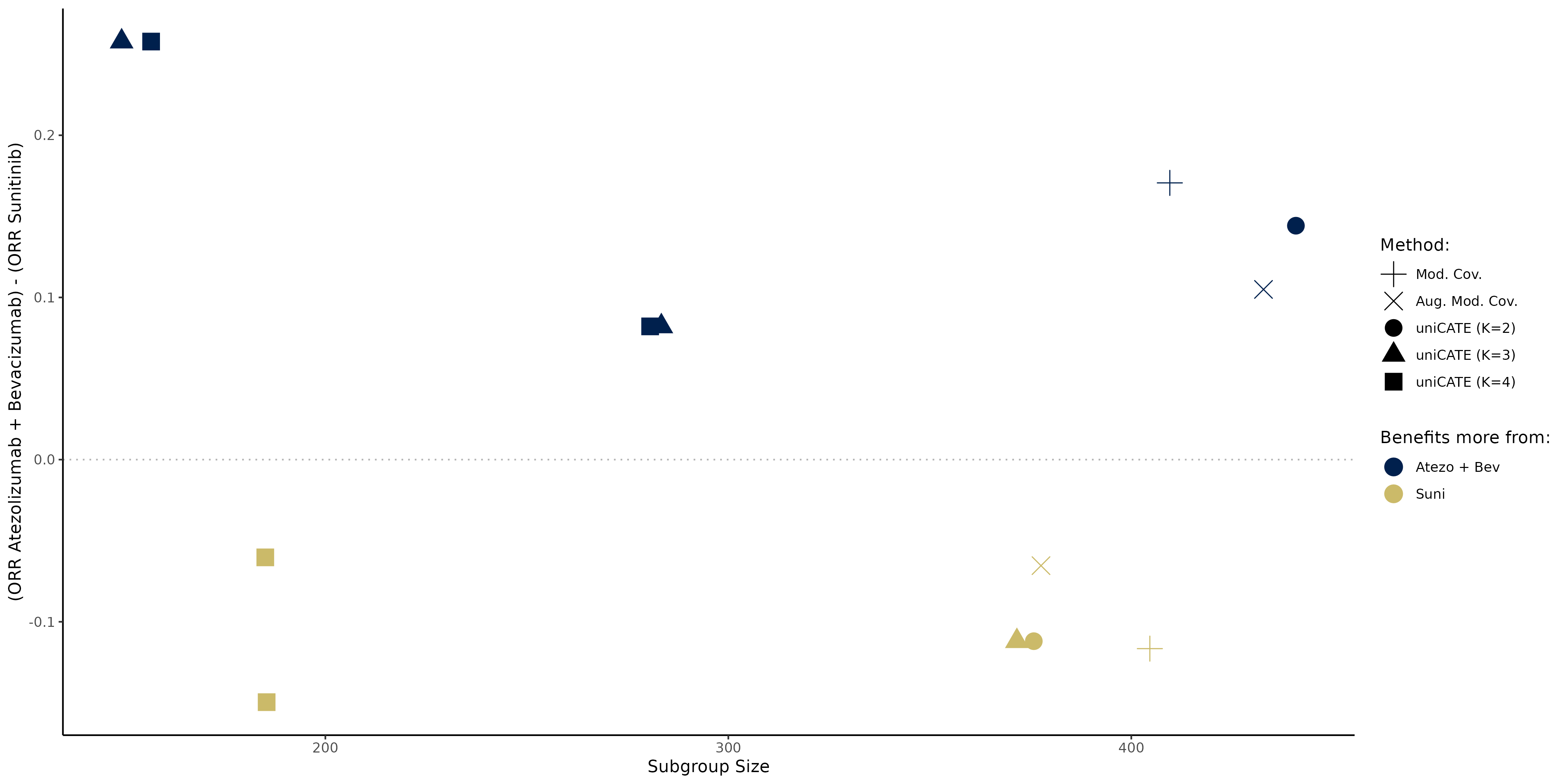}
  \caption{Comparison of the ORR across the methods' predicted subgroups in
    the IMmotion151 trial. The hierarchical clustering with complete linkage
    and Euclidean distance applied to uniCATE's predictive biomarkers was used
    to iteratively define two, three, and four clusters (K). The points are
    slightly horizontally jittered along the x-axis to avoid overplotting.}
  \label{fig:orr-comparison}
\end{figure}

It is unclear from these heatmaps alone whether these subgroups correspond to
clusters of patients that benefit more from one therapy than another. We
therefore established subgroups by performing hierarchical clustering with
complete linkage using Euclidean distance on the methods' selections of
IMmotion 151's log-transformed gene expression data. The difference in ORR was
then computed between patients receiving the atezolizumab-bevacizumab
combination and the sunitinib regiment within each of these subgroups. The
subgroups identified by the (augmented) modified covariates methods' biomarkers
had negligible differences in ORR (not shown). Instead, we used their estimated
treatment rules to predict whether each patient would benefit more from the
atezolizumab-bevacizumab combination or sunitinib, and then computed the
difference in ORR within these groups. The findings are presented in
Figure~\ref{fig:orr-comparison}.

Figure~\ref{fig:orr-comparison} evaluates the patient populations classified by
each method. When classifying patients into subgroups, the modified covariates
approach and the augmented modified covariates approach subset patients into
two groups. Ideally, one of the patient groups should produce a large, positive
ORR difference representing an increased benefit from the novel drug
combination. Figure~\ref{fig:orr-comparison} shows that when two patient
groups are of interest, uniCATE's biomarkers-defined subgroups are comparable
to the groups identified by the other two methods in terms of the effect size
and the group size. The unsupervised clustering approach used in uniCATE also
permits the definition of multiple clusters, providing a more refined
investigation of patient sub-populations. When considering three or four
clusters ($K = 3,4$), one subgroup is found to respond much better on average
to the atezolizumab-bevacizumab combination than to sunitinib. This difference
in ORR is greater than that of any subgroup defined using treatment assignment
rules.

These results suggest that uniCATE uncovered biomarkers that influence whether
mRCC patients are more likely to respond to tyrosyne kinase inhibitors alone or
in combination with immune checkpoint inhibitors. More work is necessary to
validate these findings, and to determine whether these biomarkers could form
the basis of assays that inform treatment decisions. Demonstrating that these
biomarkers are predictive of other clinical endpoints, like overall survival,
progression-free survival and safety, would make for compelling evidence.
Recovering these biomarkers in a comparison of this drug combination to the
current standard of care would be more convincing still.

\section{Discussion}\label{sec:discussion}

In this work, we demonstrate how predictive biomarker discovery, typically a
byproduct of treatment rule estimation, is better addressed as a standalone
variable importance estimation problem. We derive a novel nonparametric
estimator for a causal parameter which we argue is generally useful and
interpretable, and show that this estimator is consistent and asymptotically
linear under non-restrictive assumptions. We then verify that our proposed
procedure's asymptotic guarantees are approximately achieved across diverse
data-generating distributions in a thorough simulation study of moderate to
high-dimensional randomized control trials. Our method is then used in an
exploratory analysis of real clinical trial data, producing biologically
meaningful results that identify patient subgroups with greater treatment
effect heterogeneity than procedures not explicitly developed for predictive
biomarker discovery.

While we derive theory for uniCATE's application to observational data, we
benchmarked it exclusively in randomized control trial settings since they
constitutes our primary application area of interest. Evaluating our method in
quasi-experimental settings, however, offers an interesting avenue of future
research. Subsequent work may also explore analogous (causal) variable
importance parameters based on, for example, the relative CATE, or adapt the
univariate CATE for time-to-event outcomes. The study of other non and
semiparametric estimators of these parameters, such as one-step estimators or
targeted maximum likelihood estimators, might also prove fruitful.  Finally,
future work might assess whether treatment effect variable importance parameter
inference procedures could be coupled with novel multiple testing adjustment
approaches, like that of \citet{fithian2020}, to better account for the complex
correlation structures often found among biomarkers.

\section*{Software}

The uniCATE method is implemented in the open-source \texttt{uniCATE}
\texttt{R} software package at \url{github.com/insightsengineering/uniCATE}.
Internally, it relies on the cross-validated framework of the \texttt{origami}
\texttt{R} package \citep{coyle2018} and on Super Learning framework of the
\texttt{sl3} \texttt{R} package \citep{coyle2021sl3}. Version 0.1.0 of
\texttt{uniCATE} was used to produce the results presented in
Sections~\ref{sec:simulations} and \ref{sec:app-rct}.

\section*{Reproducibility and Data Availability}

Code to reproduce the simulation study of Section~\ref{sec:simulations} and
the analysis of Section~\ref{sec:app-rct} is available at
\url{github.com/PhilBoileau/pub_uniCATE}. 

The clinical trial data are available on the clinical study data request
platform (\url{https://vivli.org/}). The IMmotion150 clinical data was from
data cutoff date of Oct 17, 2016. The IMmotion151 clinical data was from data
cutoff date of Dec 3, 2019. For overall response rate (ORR) calculation in both
studies, a responder is defined as patients with their best confirmed overall
response by investigator of Complete Response (CR) or Partial Response (PR) per
RECIST v1.1, and non-responder otherwise.

The RNAseq data are available at:
\url{https://ega-archive.org/datasets/EGAD00001004183} and
\url{https://ega-archive.org/datasets/EGAD00001006618} The RNA-seq data was
processed by removing patients with missing baseline samples, and were
transferred to log10(CPM+1). In total of 22,997 genes were included whose
median log10(CPM+1) is larger than 0.01.

\section*{Acknowledgement}

PB gratefully acknowledges the support of the Fonds de recherche du Qu\'{e}bec
- Nature et technologies and the Natural Sciences and Engineering Research
Council of Canada. The authors thank James Duncan and Dr. Nima Hejazi for
helpful discussions about the methodology, and Dr. Romain Banchereau and Dr.
Zoe Assaf for helpful discussions about the analysis presented in
Section~\ref{sec:app-rct}. The authors would also like to thank Molly He for
preliminary simulation results that prompted this project. \textit{Conflicts of
interest: none declared.}
\section*{Appendix}

\renewcommand*{\thesubsection}{S\arabic{subsection}}
\setcounter{table}{0}
\renewcommand{\thetable}{S\arabic{table}}
\setcounter{figure}{0}
\renewcommand{\thefigure}{S\arabic{figure}}
\setcounter{equation}{0}
\renewcommand{\theequation}{S\arabic{equation}}

\subsection{Proofs}\label{sec:proofs}

\textbf{Theorem~\ref{thm:identification}: Identification and efficient
influence function.}
\begin{proof}
  
  Standard results ensure that $\Psi^F(P_X)$ is identified by $\Psi(P_0)$: By
  the law of double expectation, we find that $\mathbb{E}_{P_X}[Y^{(a)}B_j] =
  \mathbb{E}_{P_X}[\mathbb{E}_{P_X}[Y^{(a)}|W]B_j]$, and by
  A\ref{ass:unmeasured-confounding}, A\ref{ass:positivity} that
  $\mathbb{E}_{P_X}[Y^{(a)}|W] = \bar{Q}_0(a, W)$. We follow the general
  guidelines in the review of \citet{hines2021} to derive the EIF of
  $\Psi_j(P_0)$. Define the fixed distribution $P$ whose support is contained
  in the support of $P_0$. We define the parametric submodel of $P_0$ for $t
  \in [0, 1]$ as
  \begin{equation*}
    P_t = tP + (1-t)P_0.
  \end{equation*}
  Then,
  \begin{equation*}
    \begin{split}
      \text{EIF}_j(O, P_0)
      & = \frac{d}{dt}\Psi_j(P_t) \bigg|_{t=0} \\
      & = \frac{d}{dt}\frac{\mathbb{E}_{P_t}\left[\left(
        \bar{Q}_t(1, W) - \bar{Q}_t(0, W)\right)B_{j}\right]}
        {\mathbb{E}_{P_t}\left[B_{j}^2\right]} \Bigg|_{t=0} \\
      & = \frac{\frac{d}{dt}\left\{\mathbb{E}_{P_t}\left[
            \left(\bar{Q}_t(1, W) - \bar{Q}_t(0, W)\right)B_{j}
        \right]\right\} \mathbb{E}_{P_t}\left[B_{j}^2\right] -
        \mathbb{E}_{P_t}\left[
        \left(\bar{Q}_t(1, W) - \bar{Q}_t(0, W)\right)B_{j}\right]
        \frac{d}{dt}\left\{\mathbb{E}_{P_t}\left[B_{j}^2\right]\right\}}
        {\mathbb{E}_{P_t}\left[B_{j}^2\right]^2} \Bigg|_{t=0} \\
      & = \frac{1}{\mathbb{E}_{P_0}\left[B_{j}^2\right]^2}
        \left(\left(\tilde{T}(O, P_0)B_{j} -
        \mathbb{E}_{P_0}\left[\left(\bar{Q}_0(1, W) -
        \bar{Q}_0(0, W)\right)B_{j} \right]\right)
        \mathbb{E}_{P_0}\left[B_{j}^2\right] \right. \\
      & \qquad\qquad\qquad\qquad \left. -
        \mathbb{E}_{P_0}\left[
        \left(\bar{Q}_0(1, W) - \bar{Q}_0(0, W)\right)B_{j}\right]
        \left(B_{j}^2 - \mathbb{E}_{P_0}\left[B_{j}^2\right]\right)
      \right) \\
      & = \frac{\left(\tilde{T}(O, P_0) -
        \Psi_j(P_0)B_{j}\right)B_{j}}{\mathbb{E}_{P_0}\left[B_{j}^2\right]}
    \end{split}
  \end{equation*}

\end{proof}

\textbf{Corollary~\ref{cor:estimator}: Estimating equation estimator
derivation and double robustness.}

\begin{proof}
  The estimating equation estimator for the $j^\text{th}$ biomarker is given
  by:
  \begin{align*}
    0 & = \sum_{i=1}^n \text{EIF}(O_i; P_m) \\
    & = \frac{\sum_{i=1}^n \left(\tilde{T}(O_i; P_m) -
      \Psi B_{ij}\right)B_{ij}}{\sum_{i=1}^n B_{ij}^2} \\
    \implies \Psi_j^{(ee)}(P_n; P_m) & = 
      \frac{\sum_{i=1}^n \tilde{T}(O_i; P_m)B_{ij}}
      {\sum_{i=1}^n B_{ij}^2}. \\
  \end{align*}
  Then, by the Weak Law of Large Numbers,
  \begin{align*}
    \Psi_j^{(ee)}(P_n; P_m) - \Psi_j(P_0) & \rightarrow
      \frac{\mathbb{E}_{P_0}\left[\tilde{T}(O; P_m)B_{j}\right]}
      {\mathbb{E}_{P_0}\left[B_{j}^2\right]} - 
      \frac{\mathbb{E}_{P_0}\left[\left(\bar{Q}_0(1, W)-
      \bar{Q}_0(0, W)\right)B_{j}\right]}
      {\mathbb{E}_{P_0}\left[B_{j}^2\right]} \\
    & \propto \mathbb{E}_{P_0}\left[B_{j}\left(
          \frac{g_0(1, W)}{g_m(1, W)} - 1\right)\left(\bar{Q}_0(1,
        W)-\bar{Q}_m(1, W)\right) \right.\\
      & \qquad\qquad\left. - B_{j}\left(
          \frac{g_0(0, W)}{g_m(0, W)} - 1\right)\left(\bar{Q}_0(0,
        W)-\bar{Q}_m(0, W)\right)\right]. \\
  \end{align*}
  If $g_m = g_0$, then this estimator is consistent. The same is true if
  either $\lVert g_m - g_0 \rVert_{2, P_0} = o_P(1)$ or
  $\lVert \bar{Q}_m - \bar{Q}_0 \rVert_{2, P_0} = o_P(1)$.
\end{proof}

\textbf{Theorem~\ref{thm:inference}: Limiting distribution of the
estimating equation estimator.}

\begin{proof}
  Define the plug-in estimator for the univariate CATE of biomarker $j$ with
  nuisance parameters estimate using $P_m$ as $\Psi_j(P_n; P_m)$.
  Then we have through the von Mises expansion of $\Psi_j(\cdot)$ about $P_0$
  that
  \begin{equation}\label{eq:von-mises}
    \begin{split}
      \sqrt{n}\left(\Psi_j(P_n; P_m) - \Psi_j(P_0)\right)
      & = \frac{1}{\sqrt{n}}\sum_{i=1}^n \text{EIF}(O; P_0)
        - \frac{1}{\sqrt{n}}\sum_{i=1}^n \text{EIF}(O; P_m) \\
      & + \sqrt{n}\left(\mathbb{E}_{P_n} - \mathbb{E}_{P_0}\right)
      \left[\text{EIF}(O; P_m) - \text{EIF}(O; P_0)\right]
      - \sqrt{n}R(P_0, P_m).
    \end{split}
  \end{equation}
  The first term is the sum of mean-zero random variables, and so it converges
  to a Normal with variance equal to that of the EIF, scaled by $n$, as
  $n\rightarrow\infty$. The second term is the bias term that is accounted for
  by the estimating equation estimator $\Psi_j^{(ee)}(P_n;P_m)$. The
  third and fourth terms are the empirical process and remainder terms,
  respectively, and we must show that they converge to zero in probability.

  The analysis of empirical process term is identical to that of the average
  treatment effect presented in \citet{zheng2011} due to the similarity of
  these parameters. Essentially, so long as the conditional outcome regression
  and propensity score estimators converge in probability to some function
  under the L2 norm, the empirical process term is bounded in probability.
  
  We now study the remainder term:
  \begin{equation}\label{eq:remainder}
    \begin{split}
      -R(P_0, P_m)
      & = \frac{\mathbb{E}_{P_0}\left[\left(\tilde{T}(O; P_m)
        - \Psi(P_m)B_{j}\right)B_{j}\right]}
        {\mathbb{E}_{P_0}\left[B_{j}^2\right]}
      + \left(\Psi_j(P_m)-\Psi_j(P_0)\right) \\
      & = \frac{1}{\mathbb{E}_{P_0}\left[B_{j}^2\right]}
        \mathbb{E}_{P_0}\left[\tilde{T}(O; P_m)B_{j}-
           \mathbb{E}_{P_0}\left[B_{j}^2\right]\Psi_j(P_0)\right] \\
      & = \frac{1}{\mathbb{E}_{P_0}\left[B_{j}^2\right]}
      \mathbb{E}_{P_0}\left[B_{j}\left(
          T_1(O;P_m) - \bar{Q}_0(1, W)
          - T_0(O;P_m) + \bar{Q}_0(0, W)\right)
      \right] \\
      & = \frac{1}{\mathbb{E}_{P_0}\left[B_{j}^2\right]}
      \mathbb{E}_{P_0}\left[B_{j}\left(
          \frac{g_0(1, W)}{g_m(1, W)} - 1\right)\left(\bar{Q}_0(1,
        W)-\bar{Q}_m(1, W)\right) \right.\\
      & \qquad\qquad\qquad\qquad \left. - B_{j}\left(
          \frac{g_0(0, W)}{g_m(0, W)} - 1\right)\left(\bar{Q}_0(0,
        W)-\bar{Q}_m(0, W)\right)\right] \\
      & \leq \frac{1}{\mathbb{E}_{P_0}\left[B_{j}^2\right]}
      \left(\left\lvert
      \mathbb{E}_{P_0}\left[B_{j}\left(
          \frac{g_0(1, W)}{g_m(1, W)} - 1\right)\left(\bar{Q}_0(1,
        W)-\bar{Q}_m(1, W)\right) \right]\right\rvert\right.\\
      & \qquad\qquad\qquad\qquad + \left.\left\lvert\mathbb{E}_{P_0}
        \left[B_{j}\left(
          \frac{g_0(0, W)}{g_m(0, W)} - 1\right)\left(\bar{Q}_0(0,
          W)-\bar{Q}_m(0, W)\right)\right]\right\rvert\right) \\
      & \leq \frac{1}{\mathbb{E}_{P_0}\left[B_{j}^2\right]}
        \left(
        \mathbb{E}_{P_0}\left[B_{j}^2\left(
        \frac{g_0(1, W) - g_m(1, W)}{g_n(1, W)}\right)^2 \right]^{1/2}
          \mathbb{E}_{P_0}\left[\left(\bar{Q}_0(1, W)-\bar{Q}_m(1, W)
          \right)^2\right]^{1/2}\right. \\
      & \qquad\qquad\qquad\qquad + \left.
        \mathbb{E}_{P_0}\left[B_{j}^2\left(
          \frac{g_0(1, W) - g_m(1, W)}{g_m(0, W)}\right)^2\right]^{1/2}
          \mathbb{E}_{P_0}\left[\left(\bar{Q}_0(0, W)-\bar{Q}_m(0,
        W)\right)^2\right]^{1/2}\right)
    \end{split}
  \end{equation}
  If $g_0$ is known, as in a randomized control trial, then the remainder term
  is exactly zero. When neither $g_0$ or $\bar{Q}_0$ is now known, then the
  remainder term of Equation~\eqref{eq:von-mises} is $o_P(1)$ under the
  conditions of A\ref{ass:double-rate-robustness}. The conditions on
  convergence rates can be relaxed even further: The remainder term converges
  to zero in probability so long as the last line of
  Equation~\eqref{eq:remainder} is $o_P(n^{-1/2})$. That is, we may obtain our
  desired result even if, say, $\bar{Q}_m$ converges at slower rate to
  $\bar{Q}_0$ than $n^{-1/4}$ in probability so long as $g_m$ converges more
  quickly to $g_0$.
 
\end{proof}

\FloatBarrier

\subsection{Additional Simulation Results}\label{sec:add-sim-results}

\begin{figure}
  \includegraphics[width=\textwidth]{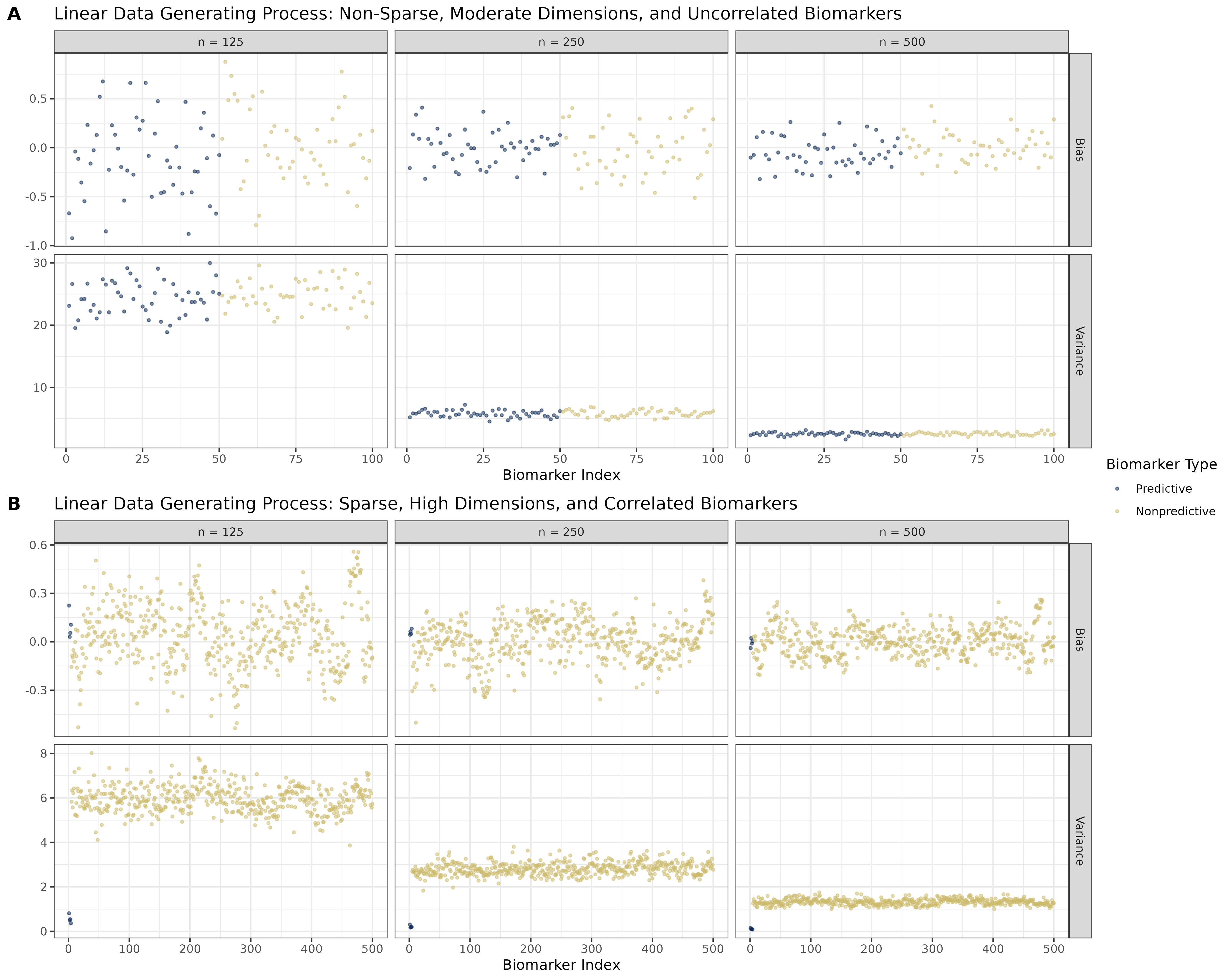}
  \caption{The empirical biases and variances of uniCATE estimates for all
  biomarkers across all simulation scenarios with a linear conditional outcome
  regression. Biomarkers coloured blue are truly predictive and those coloured
  gold are nonpredictive.}
  \label{fig:linear-bias-variance}
\end{figure}

\begin{figure}
  \includegraphics[width=\textwidth]{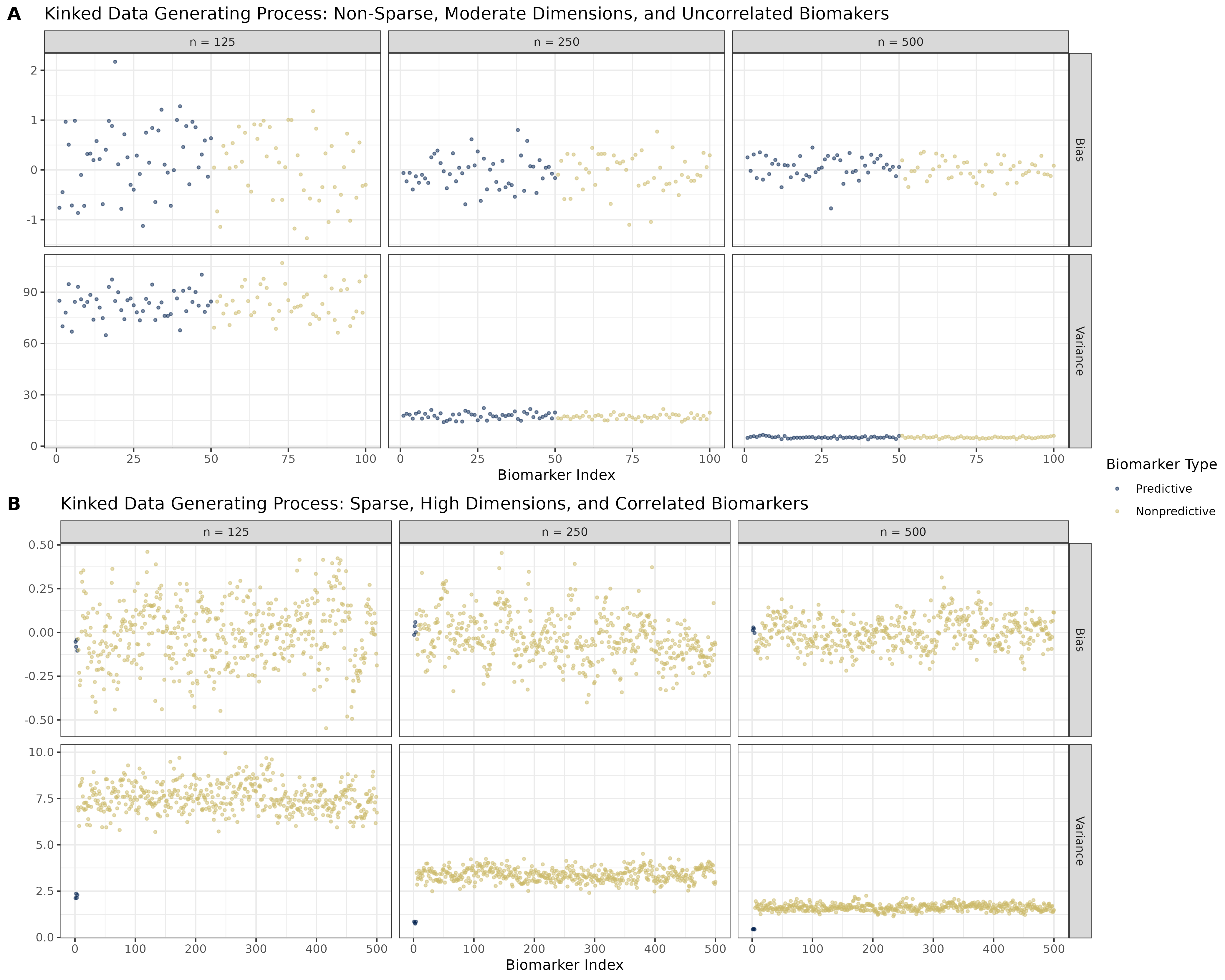}
  \caption{The empirical biases and variances of uniCATE estimates for all
  biomarkers across all simulation scenarios with a kinked conditional
  outcome regression. Biomarkers coloured blue are truly predictive and
  those coloured gold are nonpredictive.}
  \label{fig:kinked-bias-variance}
\end{figure}

\FloatBarrier

\subsection{Supporting Results from Application to IMmotion Trials}

\begin{table}[h]
  \center
  \begin{tabular}{m{10em} m{30em}}
    \hline
    Method & Predictive Biomarkers \\
    \hline
    Modified Covariates & ADCY8, CDH17, COL6A6, CSMD3, CXCL5, EEF1A2, GJB6, GRIA4, H19, IGKV1-9, KLK4, MMP3, MUC17, PZP, TCHH, TEX15, TRIM63, VIL1, WFIKKN2, XIST\\
    Augmented Modified Covariates & EEF1A2, IGKV1-9, MMP3, PZP, TEX15, TRIM63\\
    uniCATE & WFIKKN2, NMRK2, KLK1, TRIM63, IGKV1-9, HHATL, UCHL1, CLDN1, EEF1A2, C8A, KCNJ3, ITIH2, IGLV3-21, TCHH, ATP1A3, IGLL5, ENPP3, IGKV3-15, IGLC3, SAA1, TEX15, IGKV1-16, IGKV1-5, IGHG1, GRIN2A, IGHV2-5, SERPIND1, IGHV1-18, DEFB1, CYP2J2, IGHV1-24, CES3, IGKV3-11, IGLV1-40, IGHV1-2, SLC17A4, KLK4, MMP7, ANKRD36BP2, IGHV3-11, IGHV4-31, IGHV4-34, IGLV3-19, HAMP, CSMD3, PDZK1IP1, IGHG3, MUC17, ALPK2, IGLV2-14, FRAS1, DNAH11, IGHGP, SAA2, BMPER, IGLV1-47, MMP3, FOSB, HPD, SYT13, IGHV4-59, SLC38A5, IGHA1, CYP2C9, IGKC, IGLC2, PGF, IGHV3-21, H19, FCRL5, PVALB, IGHV3-74, SLC6A3, IGHV1-46, IGLV2-23, IGLV3-1, HBA1, IGLV1-44, IGKV3-20, IGKV4-1, LAMA1, IGHV3-48, IGHV5-51, IGHG2, HBA2, KNG1, IGKV1-27, IGHM, IGLV2-11, FGL1, CYP4F22, IGLV1-51\\
    \hline
  \end{tabular}
  \caption{The list of genes classified as predictive biomarkers by the considered
  methods.}
  \label{tab:predictive-biomarkers}
\end{table}

\begin{landscape}
\begin{table}
\centering
\resizebox{\linewidth}{!}{%
\begin{tabular}{m{3em} m{20em} m{3em} m{30em} m{3em} m{3em} m{3em} m{3em}}
  \hline
  Rank & Gene Set Name & Genes in Set (K) & Description & Genes in Overlap (k) & k/K & p-value & FDR q-value \\ 
  \hline
  1 & GOCC\_IMMUNOGLOBULIN\_COMPLEX & 157 & A protein complex that in its canonical form is composed of two identical immunoglobulin heavy chains and two identical immunoglobulin light chains, held together by disulfide bonds and sometimes complexed with additional proteins. An immunoglobulin complex may be embedded in the plasma membrane or present in the extracellular space, in mucosal areas or other tissues, or circulating in the blood or lymph. [GOC:add, GOC:jl, ISBN:0781765196] & 37 & 0.24 & 0 & 0 \\
  2 & GOBP\_HUMORAL\_IMMUNE \_RESPONSE\_MEDIATED\_BY \_CIRCULATING\_IMMUNOGLOBULIN & 149 & An immune response dependent upon secreted immunoglobulin. An example of this process is found in Mus musculus. [GO\_REF:0000022, GOC:add, ISBN:0781735149] & 36 & 0.24 & 0 & 0 \\
  3 & GOBP\_COMPLEMENT\_ACTIVATION & 171 & Any process involved in the activation of any of the steps of the complement cascade, which allows for the direct killing of microbes, the disposal of immune complexes, and the regulation of other immune processes; the initial steps of complement activation involve one of three pathways, the classical pathway, the alternative pathway, and the lectin pathway, all of which lead to the terminal complement pathway. [GO\_REF:0000022, GOC:add, ISBN:0781735149] & 36 & 0.21 & 0 & 0 \\
  4 & GOMF\_ANTIGEN\_BINDING & 158 & Interacting selectively and non-covalently with an antigen, any substance which is capable of inducing a specific immune response and of reacting with the products of that response, the specific antibody or specifically sensitized T-lymphocytes, or both. Binding may counteract the biological activity of the antigen. [GOC:jl, ISBN:0198506732, ISBN:0721662544] & 35 & 0.22 & 0 & 0 \\
  5 & GOBP\_B\_CELL\_MEDIATED \_IMMUNITY & 219 & Any process involved with the carrying out of an immune response by a B cell, through, for instance, the production of antibodies or cytokines, or antigen presentation to T cells. [GO\_REF:0000022, GOC:add, ISBN:0781735149] & 36 & 0.16 & 0 & 0 \\
  6 & GOBP\_HUMORAL\_IMMUNE \_RESPONSE & 373 & An immune response mediated through a body fluid. [GOC:hb, ISBN:0198506732] & 37 & 0.10 & 0 & 0 \\
  7 & GOBP\_LYMPHOCYTE \_MEDIATED\_IMMUNITY & 351 & Any process involved in the carrying out of an immune response by a lymphocyte. [GO\_REF:0000022, GOC:add, ISBN:0781735149] & 36 & 0.10 & 0 & 0 \\
  8 & GOBP\_ADAPTIVE\_IMMUNE \_RESPONSE\_BASED\_ON \_SOMATIC\_RECOMBINATION \_OF\_IMMUNE\_RECEPTORS \_BUILT\_FROM\_IMMUNOGLOBULIN \_SUPERFAMILY\_DOMAINS & 358 & An immune response mediated by lymphocytes expressing specific receptors for antigen produced through a somatic diversification process that includes somatic recombination of germline gene segments encoding immunoglobulin superfamily domains. Recombined receptors for antigen encoded by immunoglobulin superfamily domains include T cell receptors and immunoglobulins (antibodies) produced by B cells. The first encounter with antigen elicits a primary immune response that is slow and not of great magnitude. T and B cells selected by antigen become activated and undergo clonal expansion. A fraction of antigen-reactive T and B cells become memory cells, whereas others differentiate into effector cells. The memory cells generated during the primary response enable a much faster and stronger secondary immune response upon subsequent exposures to the same antigen (immunological memory). An example of this is the adaptive immune response found in Mus musculus. [GOC:add, GOC:mtg\_sensu, ISBN:0781735149, ISBN:1405196831] & 36 & 0.10 & 0 & 0 \\
  9 & GOBP\_REGULATION\_OF \_COMPLEMENT\_ACTIVATION & 114 & Any process that modulates the frequency, rate or extent of complement activation. [GOC:go\_curators] & 27 & 0.24 & 0 & 0 \\
  10 & GOBP\_PHAGOCYTOSIS & 374 & A vesicle-mediated transport process that results in the engulfment of external particulate material by phagocytes and their delivery to the lysosome. The particles are initially contained within phagocytic vacuoles (phagosomes), which then fuse with primary lysosomes to effect digestion of the particles. [ISBN:0198506732] & 35 & 0.09 & 0 & 0 \\
  \hline
\end{tabular}}
\caption{GSEA of GO terms for uniCATE's selected predictive biomarkers using IMmotion 150 data.} 
\label{tab:GOBP}
\end{table}
\end{landscape}

\begin{figure}
  \includegraphics[width=\textwidth]{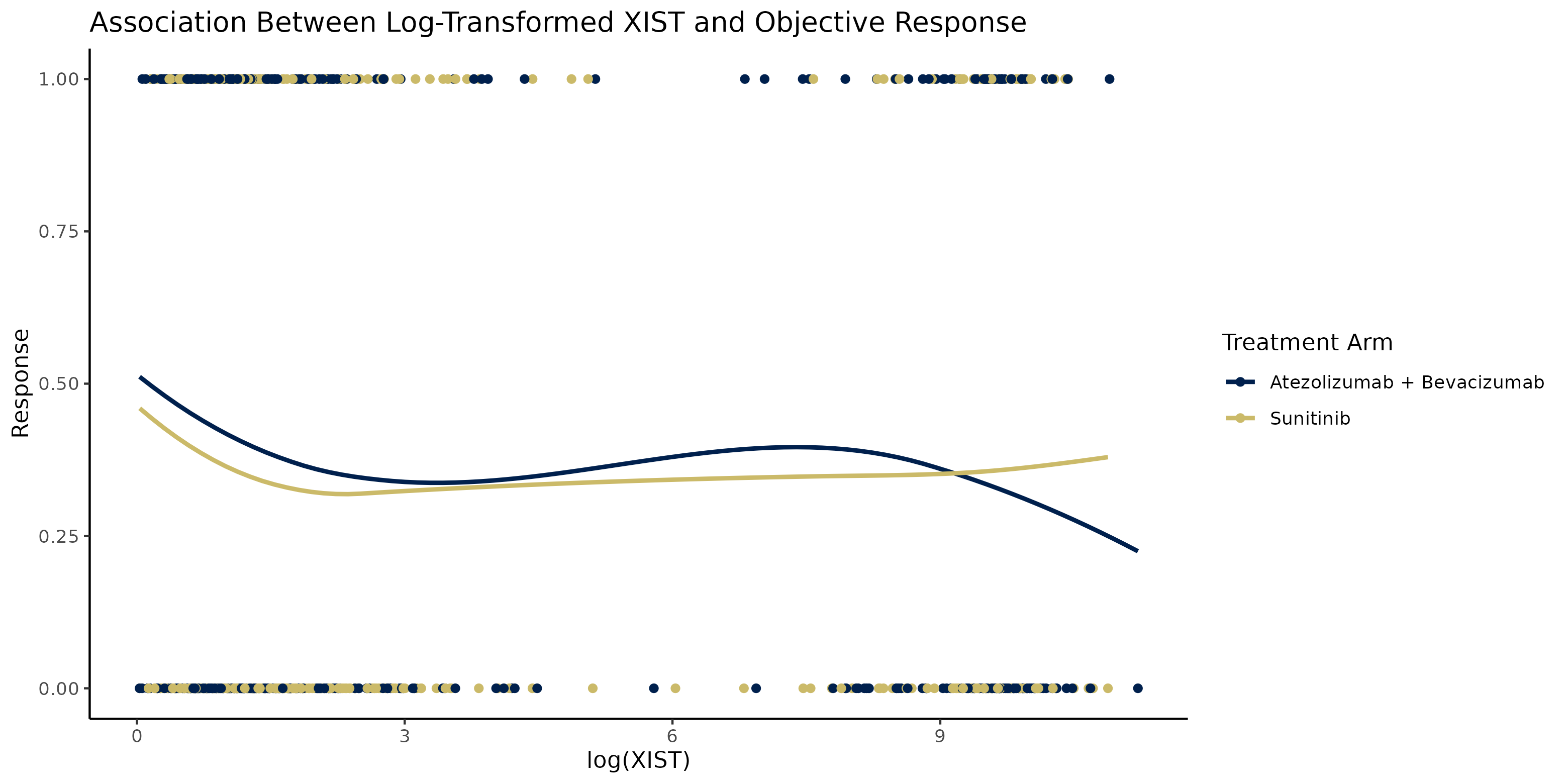}
  \caption{While the log-transformed \textit{XIST} gene expression data can be
    used to define two patient subpopulations within the IMmotion 151 study, it
    does not appear to have a strong predictive effect like the simulated
    biomarkers of Figure~\ref{fig:sketch-dgps} of the main text.}
  \label{fig:XIST}
\end{figure}

\FloatBarrier

\bibliographystyle{unsrtnat}
\bibliography{references}

\end{document}